\documentclass{article}
\PassOptionsToPackage{numbers, compress}{natbib}
\usepackage[main,final]{neurips_2026}

\usepackage[utf8]{inputenc}
\usepackage[T1]{fontenc}
\usepackage{hyperref}
\hypersetup{
	hidelinks
}
\usepackage{array}
\usepackage{url}
\usepackage{booktabs}
\usepackage{float}
\usepackage{amsfonts}
\usepackage{nicefrac}
\usepackage{microtype}
\usepackage{xcolor}
\usepackage{multirow}
\usepackage{threeparttable}
\usepackage{graphicx}
\usepackage{mathtools}
\usepackage{enumitem}
\usepackage{tcolorbox}
\usepackage{ulem}
\newcommand{\ASAP}{ASAP}

\title{\ASAP{}: Assembly-Source Aligned Pseudocode Refinement For Binary Decompilation}

\author{
  Yujian Zhuang\thanks{Equal contribution.} \\
  School of Cybersecurity\\
  Northwestern Polytechnical University\\
  Xi’an, Shaanxi 710072, China
  \And
  Dehong Gao\footnotemark[1] \\
  School of Cybersecurity\\
  Northwestern Polytechnical University\\
  Xi’an, Shaanxi 710072, China
  \AND
  Qichao Zhang \\
  School of Cybersecurity\\
  Northwestern Polytechnical University\\
  Xi’an, Shaanxi 710072, China
  \And
  Qijing Lai \\
  School of Cybersecurity\\
  Northwestern Polytechnical University\\
  Xi’an, Shaanxi 710072, China
  \AND
  Jiaxin Wang \\
  School of Cybersecurity\\
  Northwestern Polytechnical University\\
  Xi’an, Shaanxi 710072, China
  \And
  Libin Yang\thanks{
    Corresponding authors: Libin Yang
    (\href{mailto:libiny@nwpu.edu.cn}{\texttt{libiny@nwpu.edu.cn}})
    and Xiaoyan Cai
    (\href{mailto:xiaoyanc@nwpu.edu.cn}{\texttt{xiaoyanc@nwpu.edu.cn}}).
  } \\
  School of Cybersecurity\\
  Northwestern Polytechnical University\\
  Xi’an, Shaanxi 710072, China
  \AND
  Xiaoyan Cai\footnotemark[2] \\
  School of Automation\\
  Northwestern Polytechnical University\\
  Xi’an, Shaanxi 710129, China
}

\begin{document}

\maketitle

\begin{abstract}
Large language models (LLMs) are increasingly used in binary decompilation to refine the C-like pseudocode produced by traditional rule-based decompilers.
While this pseudocode is useful, it is a heuristic and lossy abstraction rather than a faithful copy of the source code.
It often contains decompiler errors, especially for aggressively optimized binaries where critical low-level details are obscured.
We present \ASAP{}, an \textbf{A}ssembly-\textbf{S}ource \textbf{A}ligned \textbf{P}seudocode refinement framework for binary decompilation.
\ASAP{} learns source-aligned assembly representations from paired source and binary functions using joint function-level and snippet-level contrastive alignment.
A Q-Former then compresses chunk-level assembly features into a fixed number of assembly tokens.
These tokens and the decompiler-produced pseudocode jointly condition the decompilation LLM.
During refinement, we use stochastic pseudocode masking and a relative assembly-advantage loss to discourage the model from relying solely on pseudocode.
On two decompilation benchmarks across multiple compiler optimization levels, \ASAP{} improves the average re-execution rate from 64.7\% to 71.9\% and the average recompilation rate from 91.6\% to 96.6\% compared with the strongest baseline.
Our approach offers both a new perspective and a practical solution to binary decompilation.
\end{abstract}

\section{Introduction}
Binary decompilation aims to recover source-level code from compiled binaries and is heavily used in reverse engineering tasks such as vulnerability detection~\cite{dinesh2020retrowrite,nagy2021breaking,lu2025automated}, malware analysis~\cite{wang2017ramblr,ming2017binsim,jang2013towards}, and closed-source program comprehension~\cite{pang2021sok,wang2026vsim,gao2024sigmadiff}.
Unlike source-to-source translation~\cite{tehranijamsaz2024coderosetta,bitan2025unipar,chen2025systematic}, decompilation is inherently underdetermined because compilation discards comments, variable names, high-level types, and many control structures.
The practical goal is typically not to reconstruct the exact original source, but to generate high-level code that is recompilable, functionally equivalent, and readable enough for human review~\cite{dramko2025fast}.

\begin{figure*}[htbp]
    \centering
    \includegraphics[width=\textwidth]{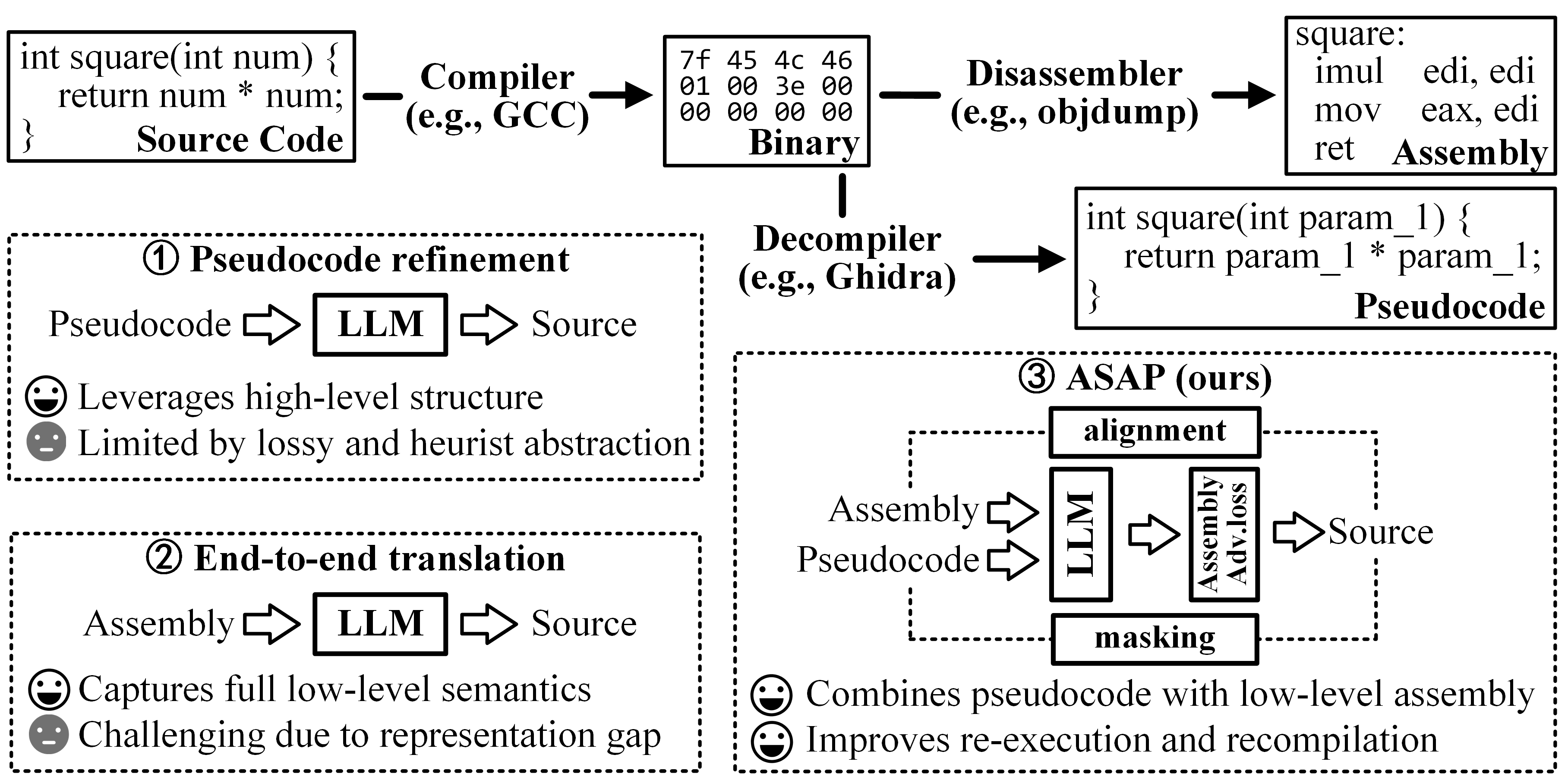}
    \caption{Comparison of our proposed \ASAP{} with existing binary decompilation approaches. The top shows the standard compilation and decompilation process from source code to binary, assembly, and decompiler-produced pseudocode. Below, three approaches are compared: \textbf{(1)} Pseudocode refinement takes decompiler output as input and improves it with an LLM; \textbf{(2)} End-to-end translation directly generates source code from assembly; \textbf{(3)} \ASAP{} augments the pseudocode refinement with assembly tokens, addressing the representation gap and attention bias.}
    \label{fig:intro}
\end{figure*}

Widely used decompilers like Ghidra~\cite{Ghidra2026} and IDA Pro~\cite{IDA2026} provide a helpful intermediate representation by translating binaries into C-like pseudocode shown in Figure~\ref{fig:intro}.
This pseudocode is more intuitive for humans than raw assembly instructions~\cite{11129273,zhuo2025beyond}; however, it remains a heuristic abstraction.
During this process, low-level details that are critical to functional equivalence may be lost or misrepresented~\cite{wiedemeier2025pylingual}.
For example, an aggressively optimized binary may encode signedness through instruction patterns such as sign extension, conditional-set instructions, and flag-dependent branches.
If the decompiler incorrectly infers an unsigned type or oversimplifies the comparison, the resulting pseudocode may appear plausible but fail on negative inputs.
Therefore, the decompiler provides a valuable yet imperfect starting point. Its output often conceals a semantically incorrect abstraction behind a readable format~\cite{de2026empirical,sirlanci2025empirical}.

Large language models (LLMs)~\cite{zhao2023survey} have shown strong potential in code-related tasks~\cite{ke2025qimeng,tang2024code,ding2024semcoder}, leading to their application in binary decompilation.
Existing approaches generally follow two directions.
The first performs pseudocode-based refinement, where an LLM rewrites decompiler-produced pseudocode into cleaner or more compilable source code~\cite{tan2024llm4decompile,tan2025sk2decompile,wong2025decllm,hu2024degpt,zhou2025fidelitygpt,cui2026pcodetrans,chen2025recopilot,wang2025context}.
This approach benefits from the structure already recovered by the decompiler and typically achieves good performance~\cite{tan2024llm4decompile}.
However, when the pseudocode contains a wrong abstraction, a refinement LLM is likely to keep the underlying errors.
The second direction translates assembly directly into source code~\cite{jiang2025nova,feng2025interactive,liu2025codeinverter,wang2025salt4decompile,liu2026decodet5}.
Although this avoids reliance on flawed pseudocode, assembly sequences are very long and their syntax differs greatly from the source-code and natural-language data on which most code LLMs are trained.
Direct assembly-to-source translation therefore remains difficult, especially for long or heavily optimized functions.

These observations suggest that pseudocode and assembly offer complementary views of the same binary.
Pseudocode provides a compact high-level hypothesis, and assembly preserves low-level operational evidence that may be missing or distorted in the decompiler output~\cite{de2026empirical,wang2025salt4decompile}.
This matches the workflow of human reverse engineers, who routinely check both pseudocode and assembly when analyzing difficult binaries~\cite{burk2022decomperson}.
However, combining both views requires solving two problems.
The first is the \textbf{representation gap} between assembly and source code. Raw assembly is much longer than source code and has a different syntax from typical LLM training data~\cite{litransforming}.
The second pertains to the \textbf{attention bias} of pseudocode.
Since pseudocode exhibits a higher degree of structural and semantic similarity to source code than assembly, the model tends to disproportionately allocate its attention to the pseudocode during the generation process (empirical analysis provided in Appendix~\ref{add:att}).
Consequently, this leads to the overshadowing or neglect of critical information embedded within the assembly itself.
The main challenge is therefore not just to append assembly to the prompt, but to encode low-level evidence compactly and train the LLM so that this evidence serves as a corrective signal rather than ignored context.

We introduce \ASAP{}, an \textbf{A}ssembly-\textbf{S}ource \textbf{A}ligned \textbf{P}seudocode refinement framework for binary decompilation.
As shown in Figure~\ref{fig:intro}, \ASAP{} keeps pseudocode as the main interface for source-level recovery but augments it with a compact learned representation of the corresponding assembly.
To bridge the representation gap, we align the semantic representation of assembly and source with semantic-parallel <assembly, source> pairs.
Our semantic alignment combines function-level and snippet-level alignment, and then query-based compression maps long assembly sequences into a fixed-size representation compatible with standard LLMs.
To alleviate the attention bias, we propose an assembly-augmented refinement fine-tuning using a pseudocode masking strategy and a detached relative assembly-advantage loss.
By stochastically masking a proportion of the pseudocode, this fine-tuning encourages the LLM to consult assembly evidence when pseudocode cues are incomplete or unreliable.
In this design, the assembly does not replace pseudocode or support standalone decompilation. Instead, it supplies targeted evidence for cases where the pseudocode hypothesis is biased or incomplete.

To validate \ASAP{}, we conduct experiments on two standard decompilation benchmarks (HumanEval~\cite{chen2021evaluating} and MBPP~\cite{austin2021program}) across four optimization levels (O0, O1, O2, and O3). The main results indicate that \ASAP{} achieves state-of-the-art performance, boosting the average re-execution rate from 64.7\% to 71.9\% and the recompilation rate from 91.6\% to 96.6\% compared to the strongest baseline.
Our contributions are threefold: \textbf{First}, our \ASAP{} introduces the alignment-then-decompilation framework, which significantly enhances the performance of binary decompilation by integrating semantic information from assembly code. \textbf{Second}, we develop a robust alignment method between assembly and source code. The pretrained assembly encoder serves as a generalized representation module that can be seamlessly transferred to various assembly-related tasks, demonstrating superior generalization capabilities. \textbf{Third}, we introduce pseudocode masking and assembly-advantage loss to effectively address the attention bias issue. To the best of our knowledge, we are the first to develop an LLM-based framework capable of integrating assembly with pseudocode for accurate binary decompilation.

\section{Related Work}
\subsection{Pseudocode-Based Refinement}
Pseudocode-based methods use decompiler-produced pseudocode as an intermediate representation for learning-based decompilation. Early work mainly focused on recovering local semantic information like variable names, types, and user-defined structures~\cite{lacomis2019dire,chen2022dirty,resym2024}. Recent LLM-based approaches further refine pseudocode at the function level, using prompting, compiler feedback, execution traces, or specialized decompilation models to improve readability, recompilability, and functional equivalence~\cite{hu2024degpt,wong2025decllm,cui2026pcodetrans,tan2024llm4decompile}. These methods benefit from the high-level structure provided by pseudocode, which also helps them remain competitive against direct assembly translation. However, pseudocode naturally carries decompiler-induced artifacts, including incorrect signedness, pointer arithmetic, and optimized control-flow idioms. A refinement LLM may therefore generate cleaner code while preserving an incorrect abstraction. \ASAP{} builds on the refinement paradigm rather than discarding it, adding compact assembly representations to enable the LLM to consult low-level operational evidence that may be missing from the pseudocode.

\subsection{End-to-End Assembly Translation}
Another line of research translates assembly instructions directly into source code. SLaDe~\cite{armengol2024slade} scales sequence-to-sequence translation with type inference-based program analysis. 
Nova~\cite{jiang2025nova} applies functionality contrastive learning across source code and assembly. It also uses optimization contrastive learning to model the ordering of optimization levels. FAE~\cite{feng2024self} uses compiler debugging information to construct fine-grained alignment data between assembly and source code for decompilation training. 
LLM4Decompile~\cite{tan2024llm4decompile} pretrains causal language models on large assembly-source datasets. Other methods improve assembly-to-source translation using interactive binary access or structural priors like control-flow graphs and source-level abstractions~\cite{feng2025interactive,liu2025codeinverter,wang2025salt4decompile}. Among these methods, DECodeT5~\cite{liu2026decodet5} is closely related because it integrates a pretrained assembly encoder into a CodeT5-based~\cite{wang2021codet5} decompilation model. However, DECodeT5 is strictly an assembly-to-source approach and uses a model primarily conditioned on assembly instructions rather than decompiler-produced pseudocode. By contrast, \ASAP{} keeps pseudocode-based refinement as the core generation interface and compresses source-aligned assembly features into a fixed-size set of assembly tokens. These tokens are not meant to encode the exhaustive information required for standalone decompilation. Instead, they serve as targeted corrective evidence for an already informative pseudocode hypothesis.

\section{Method}
\begin{figure*}[htbp]
	\centering
	\includegraphics[width=\textwidth]{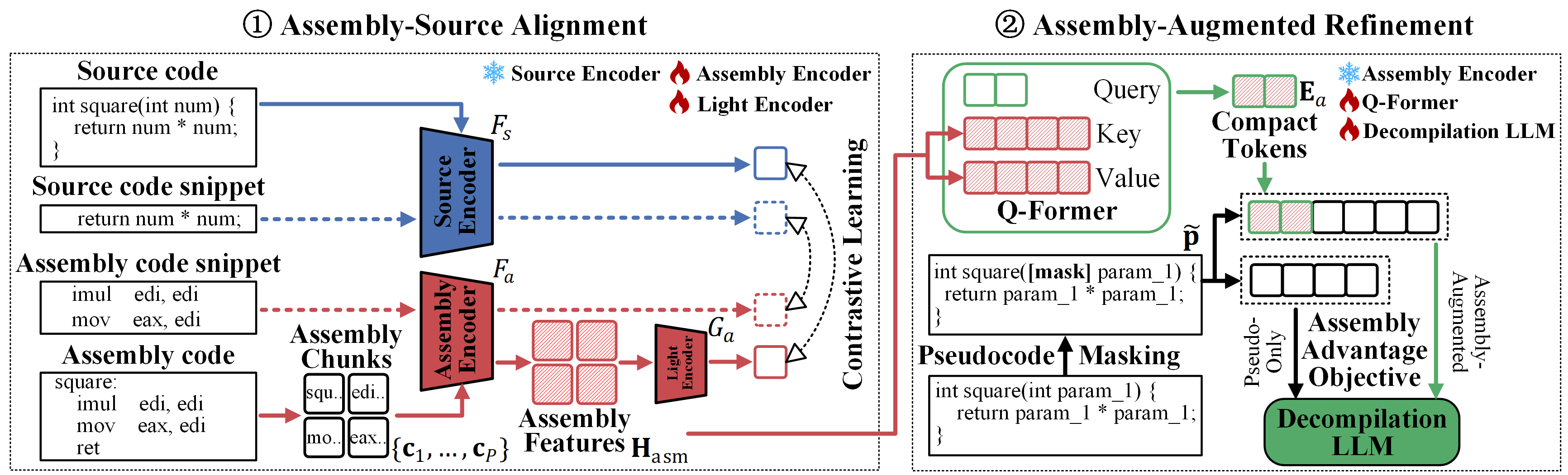}
	\caption{Our \ASAP{} framework follows a two-stage training pipeline. (1) In assembly-source alignment, an assembly encoder is trained via function-level and snippet-level contrastive learning against a frozen source encoder to align assembly and source representations. (2) In assembly-augmented refinement, the pretrained assembly encoder is frozen, while a Q-Former compresses assembly features into compact assembly tokens for the LLM, which is optimized with pseudocode masking and an assembly-advantage objective for binary decompilation.}
	\label{fig:overview}
\end{figure*}

\ASAP{} operates on compiled binary functions and uses both disassembled assembly and decompiler-produced pseudocode to recover source-level code.
As shown in Figure~\ref{fig:overview}, the framework is composed of two parts: assembly-source alignment (Section~\ref{sec:align}) and assembly-augmented refinement (Section~\ref{sec:refine}).

\subsection{Assembly-Source Alignment (ASA)}
\label{sec:align}
The syntax gap between assembly and source code makes it difficult for models to capture stable execution semantics directly from raw assembly instructions~\cite{jiang2025nova}.
To provide the decompilation LLM with useful low-level evidence, \ASAP{} first learns an assembly encoder whose representations are aligned with source-code representations.
We initialize both the assembly encoder \( F_a \) and the source encoder \( F_s \) from the same pretrained code model, but keep \( F_s \) frozen during alignment so the assembly side is optimized toward a stable source-side anchor.
The source encoder is only used to define training targets in the representation space and is discarded at inference time. Unless otherwise stated, \( \mathrm{Pool}(\cdot) \) denotes mean pooling over non-padding final-layer token states, and \( \mathrm{Norm}(\cdot) \) denotes \( \ell_2 \) normalization.

\subsubsection{Chunked Assembly Encoding}
Assembly code regularly exceeds the context window of standard code encoders (e.g., CodeT5~\cite{wang2021codet5}, UniXcoder~\cite{guo2022unixcoder}, and CodeR~\cite{litowards}).
To ensure scalable encoding, \ASAP{} partitions tokenized assembly code into \(P\) instruction-boundary chunks \( \{ \mathbf{c}_1, \dots, \mathbf{c}_P \} \), where each chunk contains at most \( M \) subword tokens.
Each chunk is processed independently through the assembly encoder, yielding a chunk representation \( \mathbf{h}_p = \mathrm{Pool}(F_a(\mathbf{c}_p)) \).
We denote the resulting chunk-level assembly features as \( \mathbf{H}_{\mathrm{asm}} = [\mathbf{h}_1; \dots; \mathbf{h}_P] \).
Keeping these features at chunk granularity preserves local evidence for the subsequent Q-Former and avoids compressing the entire function into a single vector.

For function-level contrastive learning, a single embedding is required for the entire assembly code.
We therefore introduce learnable chunk-position embeddings \( \mathbf{e}_p \) and use a light encoder \( G_a \), implemented as a shallow Transformer~\cite{vaswani2017attention}, to summarize the chunk sequence:
\begin{equation}
	\mathbf{z}_a
	=
	\mathrm{Norm}
	\left(
	\mathrm{Pool}
	\left(
	G_a \big( [\mathbf{h}_1 + \mathbf{e}_1; \dots; \mathbf{h}_P + \mathbf{e}_P] \big)
	\right)
	\right).
\end{equation}
Here, \( \mathbf{z}_a \) is the function-level assembly embedding. The corresponding source embedding \( \mathbf{z}_s \) is derived by applying the source encoder \( F_s \) to the source code \( \mathbf{y} \):
\begin{equation}
	\mathbf{z}_s
	=
	\mathrm{Norm}
	\left(
	\mathrm{Pool}
	\left(
	F_s(\mathbf{y})
	\right)
	\right).
\end{equation}
This setup allows \( F_a \) to independently produce reusable chunk-level features, while the light encoder \( G_a \) provides a global representation only when function-level contrastive learning is needed. During refinement, \ASAP{} directly uses \( \mathbf{H}_{\mathrm{asm}} \) and discards \( G_a \).

\subsubsection{Function-level Alignment}
The primary training objective aligns each assembly sequence with its corresponding source code.
A practical issue is that the same source code may be compiled under different optimization levels.
These resulting binaries should be considered functionally equivalent.
We therefore use a multi-positive contrastive loss.
For a mini-batch of \( B \) examples, let \( g(i) \) be the source-code identity of example \( i \), and let \( \mathcal{P}(i) = \{ j \mid g(j) = g(i) \} \) represent all examples in the batch compiled from the same source.
Defining \( s(\mathbf{u}, \mathbf{v}) = \exp(\mathbf{u}^\top \mathbf{v} / \tau) \), where \( \tau > 0 \) is a learnable temperature scalar, the function-level alignment loss is:
\begin{equation}
	\mathcal{L}_{\mathrm{func}}
	=
	-\frac{1}{2B}
	\sum_{i=1}^{B}
	\left[
	\log
	\frac{
		\sum_{j \in \mathcal{P}(i)} s(\mathbf{z}_{a_i}, \mathbf{z}_{s_j})
	}{
		\sum_{j=1}^{B} s(\mathbf{z}_{a_i}, \mathbf{z}_{s_j})
	}
	+
	\log
	\frac{
		\sum_{j \in \mathcal{P}(i)} s(\mathbf{z}_{s_i}, \mathbf{z}_{a_j})
	}{
		\sum_{j=1}^{B} s(\mathbf{z}_{s_i}, \mathbf{z}_{a_j})
	}
	\right].
\end{equation}
This objective is a multi-positive extension of the symmetric InfoNCE loss~\cite{oord2018representation}.
It pulls assembly and source codes sharing identical semantics closer in the joint space, reducing false negatives caused by different compiler optimization views of the same source code.

\subsubsection{Snippet-level Alignment}
Function-level alignment provides necessary coarse semantic supervision, but it may not be enough to teach the assembly encoder to reliably capture local patterns critical for accurate decompilation.
Examples include signedness, bit-width conversions, pointer arithmetic, and optimized arithmetic idioms.
To strengthen this fine-grained alignment, \ASAP{} leverages assembly-source snippet pairs built from compiler debug information (specifically, source-to-instruction line mappings).
Because compiler optimizations frequently move, merge, duplicate, or remove source statements, these pairs are treated as weak positives rather than exact statement-level correspondences.

The metadata is only used to construct weakly aligned training pairs and is never exposed to the decompilation LLM.
We discard snippets with empty mappings. Two snippets are classified as weak positives whenever their source-line metadata overlap.
Snippets originating from the same source function with overlapping metadata are excluded from one another's negative sets.
Because snippets are typically short, assembly snippets are encoded directly by \( F_a \) without chunking or the light encoder, and source snippets are encoded by \( F_s \).
We apply the same symmetric contrastive formulation as \( \mathcal{L}_{\mathrm{func}} \) and denote the resulting loss as \( \mathcal{L}_{\mathrm{snip}} \).
This biases the assembly encoder toward recovering local source-relevant cues instead of exact source-line memorization.
The final joint training objective is \( \mathcal{L}_{\mathrm{pre}} = \mathcal{L}_{\mathrm{func}} + \mathcal{L}_{\mathrm{snip}} \).
After this stage, the assembly encoder is frozen and used only to extract chunk-level assembly features for the refinement phase.

\subsection{Assembly-Augmented Refinement (AAR)}
\label{sec:refine}
After assembly-source alignment, \ASAP{} freezes the assembly encoder and proceeds to optimize the Q-Former and the decompilation LLM.
The goal is not to translate the full assembly sequence into source code. Instead, the assembly encoder yields a sequence of chunk-level features \( \mathbf{H}_{\mathrm{asm}} \), which the Q-Former compresses into a fixed-size set of assembly tokens.
The LLM then recovers source-level code from the pseudocode augmented with these assembly tokens.

\subsubsection{Assembly Feature Compression}
A simple way to combine assembly and pseudocode is direct textual concatenation.
This is inefficient because assembly code is much longer than pseudocode, and it forces the LLM to model low-level instruction syntax directly.
\ASAP{} avoids this issue by using a Q-Former~\cite{li2023blip} as a query-based compressor.
The Q-Former accepts \( K \) learnable query vectors, denoted \( \mathbf{Q} \), as queries and the chunk-level assembly features \( \mathbf{H}_{\mathrm{asm}} \) as its key-value memory. It outputs \( K \) assembly-derived vectors, which are then projected into the LLM's embedding space:
\begin{equation}
	\mathbf{R}_a = \mathrm{Q\mbox{-}Former}(\mathbf{Q}, \mathbf{H}_{\mathrm{asm}}),
	\qquad
	\mathbf{E}_a = \mathrm{Proj}(\mathbf{R}_a) \in \mathbb{R}^{K \times d_{\mathrm{LLM}}}.
\end{equation}
The resulting matrix \( \mathbf{E}_a \) is a dense assembly-token sequence that is prepended to the pseudocode token embeddings. Because \( K \) is fixed, the LLM receives a constant number of assembly tokens regardless of the original assembly length.
This design intentionally trades strict completeness for high usability. These assembly tokens do not need to encode every instruction, but are tasked with keeping the source-aligned evidence most useful for correcting the pseudocode hypothesis.

\subsubsection{Pseudocode Masking}
Decompiler-produced pseudocode supplies a high-level structure.
If the LLM always sees the complete pseudocode during training, it is likely to learn a shortcut and simply rewrite the pseudocode while ignoring the assembly tokens.
To counteract this behavior, \ASAP{} uses stochastic pseudocode masking during refinement training.
Masking acts as a regularization mechanism rather than assuming that all pseudocode errors can be readily localized through token removal.

Given pseudocode \( \mathbf{p} \), a masked variant \( \tilde{\mathbf{p}} = C_{\rho}(\mathbf{p}) \) is generated, where \( C_{\rho}(\cdot) \) denotes a corruption function parameterized by a target masking ratio \( \rho \).
The ratio \( \rho \) is sampled uniformly from \( [0, \rho_{\max}] \), and a \( \rho \) fraction of non-structural pseudocode tokens is then masked.
Structural tokens like braces, parentheses, semicolons, and control-flow keywords are preserved.
This masking strategy ensures that the decompiler-provided control structure remains visible while hiding local expressions, identifiers, constants, and operators that may require assembly evidence to recover accurately.
If the decompiler inadvertently produces an incorrect control-flow structure, masking alone cannot guarantee correction, but the assembly tokens are relied upon to supply the critical missing evidence.

\subsubsection{Assembly Advantage Objective}
While pseudocode masking weakens shortcut learning, it does not explicitly reward the decompilation LLM \(\theta \) for using the assembly tokens.
We therefore introduce a relative assembly-advantage objective.
Conditioned on identical masked pseudocode, the LLM should assign a lower generation loss to the target source when the assembly tokens are active than when they are replaced by zeros. This biases the training dynamics toward functional assembly conditioning.

Given an assembly-token sequence \( \mathbf{E} \) and the masked pseudocode \( \tilde{\mathbf{p}} \), the model generates the target source \( \mathbf{y} = (y_1, \dots, y_T) \). We define the corresponding token-averaged autoregressive negative log-likelihood \( \mathcal{N}_{\theta}(\mathbf{E}, \tilde{\mathbf{p}}, \mathbf{y}) \) as:
\begin{equation}
	\mathcal{N}_{\theta}(\mathbf{E}, \tilde{\mathbf{p}}, \mathbf{y})
	=
	-\frac{1}{T}
	\sum_{t=1}^{T}
	\log p_{\theta} (y_t \mid y_{<t}, \mathbf{E}, \tilde{\mathbf{p}}).
\end{equation}
We compute the average over target tokens to ensure the required margin is not dominated by sequence length. The assembly-augmented pathway uses \( \mathbf{E}_a \), yielding the full loss:
\begin{equation}
	\mathcal{L}_{\mathrm{full}}
	=
	\mathcal{N}_{\theta}(\mathbf{E}_a, \tilde{\mathbf{p}}, \mathbf{y}).
\end{equation}
Conversely, the pseudocode-only reference pathway replaces the assembly tokens with a zero assembly-token sequence \( \mathbf{0} \in \mathbb{R}^{K \times d_{\mathrm{LLM}}} \) of identical shape:
\begin{equation}
	\mathcal{L}_{\mathrm{pseudo}}
	=
	\mathrm{sg}
	\left[
	\mathcal{N}_{\theta}(\mathbf{0}, \tilde{\mathbf{p}}, \mathbf{y})
	\right],
\end{equation}
where \( \mathrm{sg}[\cdot] \) applies a stop-gradient operation to detach the reference loss when computing the advantage term. Gradients originating from this term thus only backpropagate through the assembly-augmented pathway. The zero assembly-token sequence does not serve as a deployment view. It acts purely as a matched-shape mathematical reference computed over the same masked pseudocode, ensuring that the comparative margin isolates and measures the utility of the assembly tokens.

We then penalize instances where the assembly-augmented pathway fails to outperform the pseudocode-only reference by a relative margin \( m \). Because both constituent losses are identically token-normalized, the margin is stably defined on comparable per-token quantities:
\begin{equation}
	m = \min(m_{\max}, \gamma \mathcal{L}_{\mathrm{pseudo}}),
\end{equation}
where \( \gamma \in (0,1) \) dictates the required relative improvement, and \( m_{\max} \ge 0 \) serves as a scale clamp to prevent gradient instability on difficult samples. The comprehensive refinement objective is formalized as:
\begin{equation}
	\mathcal{L}_{\mathrm{refine}}
	=
	\mathcal{L}_{\mathrm{full}}
	+
	\max \big( 0, \mathcal{L}_{\mathrm{full}} - \mathcal{L}_{\mathrm{pseudo}} + m \big).
\end{equation}
This term is strictly active whenever the assembly-augmented pathway fails to surpass the detached pseudocode-only reference by the designated margin. Together with pseudocode masking, it applies pointed training pressure specifically on examples where pseudocode alone poses a strong shortcut. During minibatch optimization, this objective is averaged across all examples. At inference time, \ASAP{} exclusively uses the fully trained assembly-augmented pathway with the unmasked pseudocode, bypassing any pseudocode masking or pseudocode-only reference computations.

\section{Experiments}
This section describes the experimental setup and evaluates \ASAP{} through main decompilation results, downstream binary-analysis tasks, and ablation studies.

\subsection{Settings}
\paragraph{Datasets.}
Following prior works~\cite{tan2024llm4decompile, tan2025sk2decompile,wang2025salt4decompile}, we train \ASAP{} on 100K C functions randomly sampled from ExeBench~\cite{armengol2022exebench}.
Evaluation is conducted on the C-language adaptations~\cite{tan2024llm4decompile} of HumanEval~\cite{chen2021evaluating} and MBPP~\cite{austin2021program}.
Each function is compiled using GCC v11.4.0~\cite{gcc2026} for the x86\_64 Linux platform across different optimization levels (O0--O3), yielding four binaries per function.
The corresponding assembly and pseudocode are produced using objdump v2.38~\cite{objdump2026} and Ghidra v11.0.3~\cite{Ghidra2026}, respectively.

\paragraph{Models and Optimization.}
We initialize our assembly and source encoders from UniXcoder~\cite{guo2022unixcoder}, while employing LLM4Decompile-9B~\cite{tan2024llm4decompile} as the decompilation LLM for pseudocode refinement. Detailed optimization settings are provided in Appendix~\ref{app:opt_settings}.

\paragraph{Metrics.}
We evaluate performance using the following four metrics: (1) Re-execution Rate: the percentage of generated functions that pass the official test cases; (2) Recompilation Rate: the percentage of generated functions that compile successfully without syntax errors; (3) Edit Similarity: one minus the Levenshtein edit distance between the generated function and the ground truth, normalized by the maximum length of the two sequences; (4) GPT-Judge~\cite{tan2025sk2decompile}: a readability and semantic recovery score (ranging from 1 to 5) assigned by GPT-5-mini.

\paragraph{Baselines.}
We evaluate \ASAP{} against three categories of baselines: (1) Rule-based Decompilers, including Ghidra~\cite{Ghidra2026} and IDA Pro~\cite{IDA2026}; (2) Task-Agnostic LLMs, including DeepSeek-v3.2, Grok-4-fast, and Gemini-3.1-flash; and (3) Decompilation Methods, including Ref Decompile~\cite{feng2025interactive}, LLM4Decompile~\cite{tan2024llm4decompile}, and SK2Decompile~\cite{tan2025sk2decompile}.
For inference, we set the temperature to 0.0 (i.e., greedy decoding) and the maximum output length to 4,096 tokens.
The task-agnostic LLMs are prompted with pseudocode as input.
We employ GPT-5-mini as a post-processing agent to extract the final function from the LLMs' outputs. See Appendix~\ref{app:prompt} for corresponding prompts.

\subsection{Main Results}
\begin{table*}[!htbp]
    \centering
    \footnotesize
    \setlength{\belowcaptionskip}{3pt}
    \caption{Main results of \ASAP{} and baselines across four optimization levels, averaged over HumanEval and MBPP.}
    \setlength{\tabcolsep}{1.8pt}
    \begin{threeparttable}
        \resizebox{\textwidth}{!}{
        \begin{tabular}{lccccccccccccccccccccccc}
            \hline
            \multicolumn{1}{c}{\multirow{2}{*}{Method}} & \multicolumn{5}{c}{Re-execution rate (\%)} & & \multicolumn{5}{c}{Recompilation rate (\%)} & & \multicolumn{5}{c}{Edit Similarity} & & \multicolumn{5}{c}{GPT-Judge} \\
            \cline{2-6} \cline{8-12} \cline{14-18} \cline{20-24}
            & O0 & O1 & O2 & O3 & AVG & & O0 & O1 & O2 & O3 & AVG & & O0 & O1 & O2 & O3 & AVG & & O0 & O1 & O2 & O3 & AVG \\ \hline
            \textbf{Rule-based Decompiler} & & & & & & & & & & & & & & & & & & & & & & & \\
            \hspace{0.4em} Ghidra & 35.0 & 23.4 & 22.2 & 19.0 & 24.9 & & 38.2 & 28.8 & 27.5 & 22.9 & 29.3 & & 0.29 & 0.27 & 0.26 & 0.23 & 0.26 & & 3.04 & 2.83 & 2.78 & 2.64 & 2.82 \\
            \hspace{0.4em} IDA Pro & 27.3 & 25.6 & 21.9 & 20.8 & 23.9 & & 33.1 & 30.5 & 28.7 & 27.3 & 29.9 & & 0.31 & 0.27 & 0.26 & 0.23 & 0.27 & & 3.12 & 2.88 & 2.77 & 2.70 & 2.87 \\
            \textbf{Task-Agnostic LLM} & & & & & & & & & & & & & & & & & & & & & & & \\
            \hspace{0.4em} DeepSeek-v3.2 & 69.6 & 58.7 & 58.5 & 52.3 & 59.8 & & 80.5 & 77.5 & 78.3 & 74.5 & 77.7 & & 0.43 & 0.39 & 0.37 & 0.34 & 0.38 & & 3.95 & 3.77 & 3.66 & 3.54 & 3.73 \\
            \hspace{0.4em} Grok-4-fast & 63.0 & 51.1 & 48.8 & 44.0 & 51.7 & & 71.4 & 68.1 & 66.6 & 60.5 & 66.7 & & 0.39 & 0.31 & 0.32 & 0.28 & 0.33 & & 3.77 & 3.50 & 3.53 & 3.45 & 3.56 \\
            \hspace{0.4em} Gemini-3.1-flash & 71.8 & 62.5 & 60.8 & 56.5 & 62.9 & & 81.0 & 81.6 & 81.7 & 77.8 & 80.5 & & 0.49 & 0.44 & 0.42 & 0.39 & 0.43 & & \textbf{4.37} & \textbf{4.05} & \textbf{3.99} & \textbf{3.87} & \textbf{4.07} \\
            \textbf{Decompilation Method} & & & & & & & & & & & & & & & & & & & & & & & \\
            \hspace{0.4em} Ref Decompile-6.7B & \underline{79.9} & 56.2 & 54.4 & 52.6 & 60.8 & & 91.4 & \underline{91.6} & 90.3 & 90.0 & 90.8 & & 0.62 & 0.54 & 0.53 & 0.53 & 0.56 & & 4.01 & 3.60 & 3.53 & 3.53 & 3.68 \\
            \hspace{0.4em} LLM4Decompile-9B & 72.8 & 60.6 & 59.7 & 54.9 & 62.0 & & 86.5 & 85.2 & 86.0 & 84.5 & 85.5 & & \underline{0.61} & 0.56 & \underline{0.55} & 0.53 & 0.56 & & 4.07 & 3.83 & 3.77 & 3.67 & 3.84 \\
            \hspace{0.4em} SK2Decompile-2x6.7B & 76.3 & \underline{64.3} & \underline{61.3} & \underline{56.9} & \underline{64.7} & & \underline{91.5} & \underline{91.6} & \underline{92.9} & \underline{90.6} & \underline{91.6} & & 0.60 & \underline{0.57} & 0.54 & \underline{0.55} & \underline{0.57} & & \underline{4.12} & \underline{3.89} & 3.77 & \underline{3.73} & \underline{3.88} \\
            \hspace{0.4em} \ASAP{}-9.3B & \textbf{82.2} & \textbf{73.4} & \textbf{68.5} & \textbf{63.7} & \textbf{71.9} & & \textbf{96.9} & \textbf{96.5} & \textbf{97.0} & \textbf{95.9} & \textbf{96.6} & & \textbf{0.63} & \textbf{0.59} & \textbf{0.57} & \textbf{0.56} & \textbf{0.59} & & \underline{4.12} & 3.87 & \underline{3.79} & 3.72 & 3.87 \\ \hline
        \end{tabular}
        }
        \begin{tablenotes}
            \footnotesize
            \item * \textbf{Bold} denotes the best performance. \underline{Underline} denotes the second-best performance.
        \end{tablenotes}
    \end{threeparttable}
    \label{t-results}
\end{table*}
\paragraph{Binary Decompilation.} We present the main evaluation results in Table~\ref{t-results}.
Compared with the base LLM, LLM4Decompile-9B, \ASAP{} improves all four metrics, with substantial gains in re-execution and recompilation rates (from 62.0\% to 71.9\% and 85.5\% to 96.6\%, respectively).
These results demonstrate the effectiveness of augmenting pseudocode refinement with source-aligned assembly evidence, leading to superior functional equivalence and syntactic correctness.
Compared with the strongest decompilation method, SK2Decompile, \ASAP{} is only slightly lower on GPT-Judge.
This is because SK2Decompile adopts a two-stage ``skeleton-to-skin'' pipeline with a dedicated identifier naming component, which improves the readability of recovered identifiers.
We also observe that Gemini-3.1-flash leads on GPT-Judge because it tends to generate long, highly readable identifier names.
These names often differ lexically from the original variables, making Gemini-3.1-flash underperform decompilation methods in edit similarity.
Overall, \ASAP{} exhibits superior performance in binary decompilation, notably outperforming other baselines on re-execution and recompilation rates.
We also provide additional results on BringUpBench~\cite{austin_bringup_bench} (Appendix~\ref{app:bringupbench}) and comparisons with frontier models and other baselines~\cite{basque2024sailr,feng2024self,wong2025decllm} (Appendix~\ref{app:additional_baselines}).

\begin{table*}[!htbp]
    \centering
    \footnotesize
    \setlength{\belowcaptionskip}{3pt}
    \caption{Downstream binary analysis results using different decompilation outputs as input.}
    \label{t-downstream}
    \newcolumntype{S}{>{\centering\arraybackslash}p{4pt}}
    \setlength{\tabcolsep}{4pt}
    \begin{threeparttable}
        \begin{tabular}{lcScccScSc}
            \hline
            \multicolumn{1}{c}{\multirow{2}{*}{Method}} & \multicolumn{1}{c}{SR} & & \multicolumn{3}{c}{BCS} & & \multicolumn{1}{c}{AC} & & \multicolumn{1}{c}{\multirow{2}{*}{AVG}} \\
            \cline{2-2} \cline{4-6} \cline{8-8}
            & Rouge-L & & BLEU & METEOR & Rouge-L & & ACC & & \\ \hline
            \hspace{0.4em} Qwen3-Coder-30B & 52.86 & & 4.63 & 23.32 & 20.39 & & 76.47 & & 35.53 \\
            \hspace{1.2em} + Gemini-3.1-flash & 61.88 & & \textbf{5.14} & 23.06 & \textbf{21.67} & & \textbf{80.88} & & \textbf{38.53} \\
            \hspace{1.2em} + SK2Decompile & \underline{63.50} & & 4.60 & \textbf{24.09} & 21.14 & & \underline{77.94} & & 38.25 \\
            \hspace{1.2em} + \ASAP{} & \textbf{63.54} & & \underline{4.94} & \underline{23.60} & \underline{21.49} & & \underline{77.94} & & \underline{38.30} \\ \hline
            \hspace{0.4em} DeepSeek-Coder-V2-16B & 39.67 & & 3.81 & 18.81 & 21.15 & & 47.06 & & 26.10 \\
            \hspace{1.2em} + Gemini-3.1-flash & \underline{62.78} & & \underline{4.39} & 19.94 & \underline{23.55} & & \underline{57.35} & & \underline{33.60} \\
            \hspace{1.2em} + SK2Decompile & 62.66 & & 4.36 & \underline{20.64} & 23.30 & & 55.88 & & 33.37 \\
            \hspace{1.2em} + \ASAP{} & \textbf{64.73} & & \textbf{5.18} & \textbf{22.20} & \textbf{23.79} & & \textbf{60.29} & & \textbf{35.24} \\ \hline
            \hspace{0.4em} CodeLlama-13B & 39.66 & & 5.51 & 26.32 & 21.94 & & 51.47 & & 28.98 \\
            \hspace{1.2em} + Gemini-3.1-flash & 61.79 & & \textbf{7.46} & \textbf{29.65} & \textbf{24.97} & & \underline{60.29} & & \underline{36.83} \\
            \hspace{1.2em} + SK2Decompile & \underline{61.87} & & 6.88 & \underline{28.76} & \underline{24.62} & & 58.82 & & 36.19 \\
            \hspace{1.2em} + \ASAP{} & \textbf{63.63} & & \underline{6.94} & 28.42 & 24.41 & & \textbf{61.76} & & \textbf{37.03} \\ \hline
        \end{tabular}
        \begin{tablenotes}
            \footnotesize
            \item * "+" denotes replacing the pseudocode with the corresponding decompilation output under the same downstream LLM and task prompt.
        \end{tablenotes}
    \end{threeparttable}
\end{table*}

\paragraph{Downstream Binary Analysis.}
Many reverse-engineering tasks operate on decompiler-produced pseudocode.
We therefore evaluate whether different decompilation outputs better support three pseudocode-based tasks in BinMetric~\cite{shang2025binmetric}: Signature Recovery (SR), Binary Code Summarization (BCS), and Algorithm Classification (AC).
SR and BCS evaluate semantic comprehension, while AC evaluates logical analysis.
We use Qwen3-Coder-30B~\cite{yang2025qwen3}, DeepSeek-Coder-V2-16B~\cite{zhu2024deepseek}, and CodeLlama-13B~\cite{roziere2023code} as downstream LLMs, replacing their original pseudocode input with outputs from Gemini-3.1-flash, SK2Decompile, and \ASAP{} under the same task prompts.
As shown in Table~\ref{t-downstream}, \ASAP{} achieves the best average downstream performance on two of the three LLMs, while remaining highly competitive on Qwen3-Coder-30B.
\ASAP{} exhibits the clearest and most consistent gains on SR, dominating this task across all baseline LLMs, which indicates its superior preservation of functional equivalence.
Gemini-3.1-flash benefits BCS by producing semantically informative identifiers that expose functional intent.
Meanwhile, although Gemini achieves the highest AC score on Qwen3-Coder-30B, \ASAP{} secures the leading AC performance on both DeepSeek-Coder and CodeLlama.
Overall, these results indicate that \ASAP{} produces high-level representations of the binary that are more useful for subsequent analysis tasks.

\subsection{Ablation Study}
\begin{table*}[!htbp]
    \centering
    \footnotesize
    \setlength{\belowcaptionskip}{4pt}
    \caption{\ASAP{} and its ablation studies.}
    \setlength{\tabcolsep}{2pt}
    \begin{threeparttable}
        \begin{tabular}{lccccccccccc}
            \hline
            \multicolumn{1}{c}{\multirow{2}{*}{Ablation study}} & \multicolumn{5}{c}{Re-execution rate (\%)} & & \multicolumn{5}{c}{Recompilation rate (\%)} \\
            \cline{2-6} \cline{8-12}
            & O0 & O1 & O2 & O3 & AVG & & O0 & O1 & O2 & O3 & AVG \\ \hline
            \textbf{Benefits from ASA} & & & & & & & & & & & \\
            \hspace{0.4em} Raw-Asm & 80.36 & 67.13 & 63.32 & 60.06 & 67.71 & & 95.59 & 95.23 & 94.77 & 94.46 & 95.01 \\
            \hspace{0.4em} w/o CL & 77.54 & 67.89 & 63.03 & 60.89 & 67.34 & & 95.03 & 95.29 & 93.90 & 93.14 & 94.34 \\
            \hspace{0.4em} w/o SLA & 79.91 & 69.83 & 65.77 & 60.61 & 69.03 & & 95.74 & 95.90 & 95.29 & 94.62 & 95.39 \\
            \hspace{0.4em} w/o FLA & 80.26 & 69.94 & 64.35 & 60.98 & 68.88 & & 96.00 & 96.05 & 94.37 & 94.17 & 95.15 \\ \hline
            \textbf{Benefits from AAR} & & & & & & & & & & & \\
            \hspace{0.4em} w/o AFC & 80.11 & 69.28 & 63.74 & 62.11 & 68.81 & & 96.11 & 96.97 & 97.18 & 96.51 & 96.70 \\
            \hspace{0.4em} w/o AAO & 78.22 & 67.43 & 64.24 & 59.14 & 67.26 & & 94.98 & 95.23 & 94.77 & 94.16 & 94.79 \\
            \hspace{0.4em} w/o PM & 79.91 & 70.50 & 64.24 & 61.89 & 69.13 & & 94.31 & 95.08 & 94.62 & 94.06 & 94.52 \\
            \hspace{0.4em} FL & 75.01 & 64.33 & 59.61 & 56.34 & 63.82 & & 88.84 & 89.80 & 90.02 & 89.06 & 89.43 \\ \hline
            \hspace{0.4em} \ASAP{} (Full) & \textbf{82.15} & \textbf{73.36} & \textbf{68.52} & \textbf{63.68} & \textbf{71.92} & & \textbf{96.92} & \textbf{96.46} & \textbf{97.02} & \textbf{95.86} & \textbf{96.57} \\ \hline
        \end{tabular}
    \end{threeparttable}
    \label{t-ablation}
\end{table*}
\paragraph{Benefits from ASA.}
As shown in Table~\ref{t-ablation}, we compare the proposed framework with four variants that remove or simplify the alignment stage: using raw assembly text together with pseudocode instead of the aligned assembly module (Raw-Asm), removing contrastive alignment training (w/o CL), removing snippet-level alignment (w/o SLA), and removing function-level alignment (w/o FLA).
Raw-Asm and w/o CL show similar average re-execution rates, 67.71\% and 67.34\%, respectively.
This suggests that an unaligned assembly encoder provides roughly the same usable evidence as directly fine-tuning the LLM with raw assembly text.
Both variants expose low-level information to the LLM, but neither organizes it into source-aligned evidence that can be reliably used during refinement.

w/o SLA and w/o FLA also perform similarly, with average re-execution rates of 69.03\% and 68.88\%.
Removing either objective causes a clear drop from the baseline, indicating that local and global alignment make comparable contributions to final performance.
Their recompilation rates remain close to \ASAP{}, suggesting that both losses mainly affect semantic recovery rather than syntax generation.
The two objectives therefore provide supervision at different granularities, and removing either one weakens the semantic usefulness of the learned assembly representation.

\paragraph{Benefits from AAR.}
Table~\ref{t-ablation} also evaluates four AAR ablations: removing assembly feature compression (w/o AFC), removing the assembly-advantage objective (w/o AAO), removing pseudocode masking (w/o PM), and freezing the decompilation LLM (FL). FL gives the largest drop, reducing the average re-execution rate to 63.82\%, close to LLM4Decompile. This is expected because the assembly tokens are continuous Q-Former outputs rather than native LLM token embeddings. When the LLM is frozen, the assembly--LLM interface cannot be adapted, so the prefix is only weakly interpretable while the prompt and pseudocode remain sufficient for pseudocode-based refinement.

w/o AAO gives the next largest drop in re-execution, but its recompilation rate remains much higher than FL. This suggests that the LLM still learns to emit valid code, yet loses semantic corrections that depend on assembly evidence. w/o AFC shows a different pattern: the decline is mild at O0 and O3, but larger at O1 and O2. At O0, pseudocode already preserves most source-level logic. At O3, strong low-level cues can still be exploited from raw assembly features. In contrast, intermediate optimizations require more selective retrieval of subtle assembly signals. w/o PM degrades performance more uniformly, indicating that pseudocode masking is a general training bias across optimization levels. Overall, these results show that each component contributes to the final performance.

\paragraph{Hyperparameters Analysis}
We analyze the sensitivity of \ASAP{} to three key hyperparameters: the number of assembly tokens $K$, the maximum pseudocode masking ratio $\rho_{\max}$, and the relative improvement ratio $\gamma$.
The experimental results and analysis are illustrated in Appendix~\ref{app:hyperparameters_analysis}.

\section{Limitations}
\label{app:limitations}
\paragraph{Functional Equivalence.}
\ASAP{} employs official test cases to identify functional inequivalence between the original source code and decompiled code.
However, formally verifying equivalence for arbitrary programs remains undecidable in light of Rice's theorem~\cite{rice1953classes}, so test-based re-execution should be interpreted as strong empirical evidence rather than a proof.
Although \ASAP{} primarily targets re-execution, integrating more fast and robust equivalence checking techniques~\cite{dramko2025fast} offers a promising path toward stronger functional validation.

\paragraph{Readability.}
\ASAP{} mainly optimizes functional recovery and recompilability, and does not include a dedicated module for recovering human-written identifiers or comments.
Since compilation often removes such information, readability-oriented evaluation may favor systems with specialized naming or rewriting components ~\cite{tan2025sk2decompile}.
Combining \ASAP{} with identifier, type, and comment recovery ~\cite{lacomis2019dire,chen2022dirty,resym2024} is an orthogonal direction that may further improve human usability without weakening its decompilation ability.

\paragraph{Model Scale.}
Prior work \cite{tan2024llm4decompile} suggests that increasing the scale of training corpus and decompilation LLMs can improve generation quality.
Due to resource constraints, our experiments do not exhaustively train \ASAP{} on frontier-scale LLMs, and the reported results should be viewed as validating the assembly-source alignment and assembly-augmented refinement principle rather than saturating its upper bound.
Because \ASAP{} provides a fixed number of assembly tokens and uses model-agnostic training objectives, we believe the framework can complement larger LLMs and guide future scaling efforts.

\paragraph{Program Structure.}
\ASAP{} uses program structure indirectly through Ghidra pseudocode, preserving structural tokens and control-flow keywords during masking. However, its assembly encoder processes instruction chunks with positional embeddings without explicitly modeling control-flow or data-flow edges between basic blocks. Future work will explore incorporating these edges into the assembly encoder or Q-Former to aggregate evidence from related basic blocks, and combining \ASAP{} with a skeleton-to-skin pipeline~\cite{tan2025sk2decompile} that uses assembly evidence for structure recovery and handles identifier naming separately.

\section{Conclusion}
We introduced \ASAP{}, an assembly-augmented pseudocode refinement framework for binary decompilation.
\ASAP{} bridges the gap between pure pseudocode refinement and direct assembly-to-source translation.
By aligning assembly representations with source code and compressing them into a fixed number of assembly tokens, \ASAP{} provides decompilation LLMs with targeted low-level evidence without overwhelming the context window.
Combined with pseudocode masking and an assembly-advantage training objective, our approach prevents the model from passively adopting decompiler errors.
Empirical results across multiple optimization levels demonstrate that \ASAP{} significantly outperforms existing methods in both recompilation and re-execution rates.
Future work will explore applying this framework to obfuscated binaries and expanding the training phase to support multiple instruction set architectures.

\section*{Acknowledgments}
This work was supported in part by the National Natural Science Foundation of China under Grants 62372380, U22B2036, Key Research and Development Plan of Shaanxi Province under Grant 2024GX-ZDCYL-01-05, Natural Science Basic Research Program of Shaanxi under Grants 2024JC-YBMS-513, the CCF-NSFOCUS ‘Kunpeng’ Research Fund under Grant CCF-NSFOCUS2025003, and Huawei Research Project under Grant TC20250730001.

\bibliographystyle{plain}
\bibliography{refs}
\newpage
\appendix
\section{Experimental Details}

\subsection{Optimization Settings}
\label{app:opt_settings}
\paragraph{Assembly-Source Alignment.}
Both the source and assembly encoders are initialized from UniXcoder (unixcoder-base-nine\footnote{\url{https://huggingface.co/microsoft/unixcoder-base-nine}}), using the same architecture and initial weights.
The light encoder is implemented as a two-layer Transformer using the Huggingface Transformers\footnote{\url{https://github.com/huggingface/transformers}} v4.57.6, with 8 attention heads in each layer.
We also apply PatchDropout~\cite{li2023scaling} with a drop rate of 0.5, which randomly drops 50\% of assembly chunks during training.
Together with BF16 mixed-precision training, PatchDropout reduces memory usage and accelerates training.
The trainable modules in this stage are optimized for one epoch using AdamW \cite{loshchilov2017decoupled} with a learning rate of $1\times 10^{-5}$ and a cosine learning-rate schedule.
We set the number of chunks to $P=64$ and cap each chunk at $M=256$ subword tokens.

\paragraph{Assembly-Augmented Refinement.}
The Q-Former follows the default configuration in LAVIS v1.0.2~\cite{li-etal-2023-lavis}. To reduce memory overhead, we use BF16 mixed-precision training and apply Low-Rank Adaptation (LoRA)~\cite{hu2022lora} only to the decompilation LLM. The LoRA rank is set to 16, alpha to 32, and the target modules include the embedding layer, lm head, and all projection layers. The trainable modules in this stage are optimized for one epoch using AdamW \cite{loshchilov2017decoupled} with a learning rate of $2\times 10^{-5}$ and a cosine learning-rate schedule.

\subsection{Prompt Templates}
\label{app:prompt}
\begin{tcolorbox}[
    colback=gray!5!white,
    colframe=gray!75!black,
    title={Prompt Template for Refinement},
    fontupper=\raggedright,
    arc=2mm,
]
    \small
    This is the decompiled code:
    \\\textcolor{blue}{\ttfamily \{DECOMPILED\_CODE\}}
    \\What is the source code?
\end{tcolorbox}

\begin{tcolorbox}[
    colback=gray!5!white,
    colframe=gray!75!black,
    title={Prompt Template for Code Extractor},
    fontupper=\raggedright,
    arc=2mm,
]
    \small
    You are a C code extractor. Your task is to analyze the provided text and perform the following actions, strictly in this order:

    1. Identify all complete C function definitions.

    2. Select and output the code for the complete C function named \textcolor{blue}{\ttfamily \{FUNCTION\_NAME\}}. If multiple functions share this name, select the last occurrence. If no function named \textcolor{blue}{\ttfamily \{FUNCTION\_NAME\}} is found, select the last complete function in the text and replace its name with \textcolor{blue}{\ttfamily \{FUNCTION\_NAME\}}.

    3. If no complete function is found, but partial code exists, output only that partial code.

    4. If no {language} code is identifiable, output an empty response.

    5. Your output MUST contain only the selected function, enclosed within <function></function> tags, with no additional text or explanation.

    [Start of Text]

    \textcolor{blue}{\ttfamily \{LLM\_OUTPUT\}}

    [End of Text]
\end{tcolorbox}

\section{Additional Experiments}
\label{app:addition}
\subsection{Hyperparameters Analysis}
\label{app:hyperparameters_analysis}
\begin{table*}[!htbp]
    \centering
    \footnotesize
    \setlength{\belowcaptionskip}{4pt}
    \caption{Effect of hyperparameters. When varying a specific hyperparameter, the others are fixed to the default configuration of \ASAP{} ($K=32$, $\rho_{\max}=0.1$, $\gamma=0.1$).}
    \setlength{\tabcolsep}{2pt}
    \begin{threeparttable}
        \begin{tabular}{lccccccccccc}
            \hline
            \multicolumn{1}{c}{\multirow{2}{*}{Hyperparameters}} & \multicolumn{5}{c}{Re-execution rate (\%)} & & \multicolumn{5}{c}{Recompilation rate (\%)} \\
            \cline{2-6} \cline{8-12}
            & O0 & O1 & O2 & O3 & AVG & & O0 & O1 & O2 & O3 & AVG \\ \hline
            \hspace{0.6em} $K=8$ & 78.89 & 68.26 & 63.13 & 62.55 & 68.21 & & 94.82 & 94.98 & 95.80 & 94.57 & 95.04 \\
            \hspace{0.6em} $K=16$ & 78.88 & 69.38 & 61.96 & 59.10 & 67.33 & & 95.03 & 95.29 & 94.07 & 93.96 & 94.59 \\
            \hspace{0.6em} $K=64$ & 80.06 & 67.80 & 63.23 & 61.39 & 68.12 & & 95.64 & 95.70 & 95.04 & 94.12 & 95.13 \\ \hline
            \hspace{0.6em} $\rho_{\max}=0.3$ & 80.06 & 70.40 & 65.11 & 62.30 & 69.47 & & 95.44 & 96.21 & 95.55 & 94.57 & 95.44 \\
            \hspace{0.6em} $\rho_{\max}=0.5$ & 78.88 & 67.68 & 63.11 & 58.78 & 67.11 & & 95.64 & 96.15 & 95.90 & 95.34 & 95.76 \\
            \hspace{0.6em} $\rho_{\max}=0.8$ & 76.38 & 63.36 & 59.96 & 56.85 & 64.14 & & 94.92 & 93.80 & 94.06 & 92.44 & 93.81 \\ \hline
            \hspace{0.6em} $\gamma=0.3$ & 79.14 & 68.25 & 65.05 & 58.79 & 67.81 & & 95.59 & 95.18 & 94.57 & 93.66 & 94.75 \\
            \hspace{0.6em} $\gamma=0.5$ & 80.57 & 72.08 & 65.72 & \textbf{63.72} & 70.53 & & 96.26 & \textbf{96.77} & 96.21 & 95.29 & 96.13 \\
            \hspace{0.6em} $\gamma=0.8$ & 79.66 & 68.31 & 63.02 & 60.12 & 67.78 & & 96.41 & 96.31 & 96.67 & 95.65 & 96.26 \\ \hline
            \hspace{0.6em} \ASAP{} & \textbf{82.15} & \textbf{73.36} & \textbf{68.52} & 63.68 & \textbf{71.92} & & \textbf{96.92} & 96.46 & \textbf{97.02} & \textbf{95.86} & \textbf{96.57} \\ \hline
        \end{tabular}
    \end{threeparttable}
    \label{t-hyperparam}
\end{table*}
As shown in Table~\ref{t-hyperparam}, we analyze the sensitivity of \ASAP{} to three key hyperparameters: the number of assembly tokens $K$, the maximum pseudocode masking ratio $\rho_{\max}$, and the relative improvement ratio $\gamma$. We vary one hyperparameter at a time around the default setting and report results for $K\in\{8,16,64\}$, $\rho_{\max}\in\{0.3,0.5,0.8\}$, and $\gamma\in\{0.3,0.5,0.8\}$.

For $K$, the three settings give close average re-execution rates, and increasing the number of assembly tokens does not bring a monotonic gain. This suggests that the Q-Former can extract useful assembly evidence with a moderate token budget, while additional queries may introduce redundant information. The $K=64$ setting is slightly stronger than w/o CL in Table~\ref{t-ablation}. This is reasonable because we set the number of assembly chunks to 64, so $K=64$ reduces the compression pressure and additionally introduces the Q-Former aggregation module.

For $\rho_{\max}$, larger masking ratios reduce re-execution, especially at $\rho_{\max}=0.8$. This indicates that pseudocode masking should remain mild, so that it weakens shortcut learning without removing too much high-level structure. For $\gamma$, $\gamma=0.5$ gives the best re-execution among the tested values, whereas $\gamma=0.8$ slightly improves recompilation but lowers re-execution. The default configuration $(K,\rho_{\max},\gamma)=(32,0.1,0.1)$ performs best overall, suggesting that \ASAP{} benefits from a moderate number of assembly tokens, mild pseudocode masking, and a mild assembly-advantage penalty.

\subsection{Attention Bias Analysis}
\label{add:att}
A key question in assembly-augmented refinement is whether the decompilation LLM actively consults the assembly tokens or merely exhibits an attention bias toward the structurally easier pseudocode.
To investigate this, we operationalize attention bias as the model's behavioral over-reliance on the pseudocode view during generation.
We introduce causal view interventions to isolate and quantify the functional contribution of both the assembly and pseudocode inputs.

We evaluate the proposed framework under four inference conditions. The \textbf{Full} condition provides the standard input: correct assembly tokens and unmodified pseudocode. The \textbf{Zero-Asm} and \textbf{Shuffled-Asm} conditions corrupt the assembly view by replacing the tokens with zero vectors and tokens from a randomly sampled function, respectively, while leaving the pseudocode intact. Conversely, the \textbf{Masked-Pseudo} condition preserves the correct assembly tokens but stochastically masks pseudocode tokens at a ratio of $\rho_{\max}=0.1$. This ratio follows the hyperparameter analysis and avoids turning the inference condition into assembly-dominant decompilation.

We report the re-execution and recompilation rates for each condition. The true utility of the assembly view is reflected in the absolute performance drop when assembly tokens are corrupted. If the model suffers from pseudocode attention bias, zeroing or shuffling the assembly would yield negligible drops, whereas masking the pseudocode would critically impair generation. Conversely, substantial degradation under the Zero-Asm and Shuffled-Asm conditions indicates that the LLM functionally depends on the assembly evidence.

\begin{table*}[!htbp]
    \centering
    \footnotesize
    \setlength{\belowcaptionskip}{4pt}
    \caption{Causal view intervention performance across different optimization levels. Zero-Asm and Shuffled-Asm test the functional reliance on assembly tokens. Larger performance drops in the AVG columns indicate that the model effectively mitigates the attention bias toward pseudocode.}
    \setlength{\tabcolsep}{3.5pt}
    \begin{tabular}{lccccccccccc}
        \hline
        \multicolumn{1}{c}{\multirow{2}{*}{Method}} & \multicolumn{5}{c}{Re-execution rate (\%)} & & \multicolumn{5}{c}{Recompilation rate (\%)} \\
        \cline{2-6} \cline{8-12}
        & O0 & O1 & O2 & O3 & AVG & & O0 & O1 & O2 & O3 & AVG \\ \hline
        Full          & 82.2 & 73.4 & 68.5 & 63.7 & 71.9 & & 96.9 & 96.5 & 97.0 & 95.9 & 96.6 \\
        Zero-Asm      & 74.5 & 64.0 & 62.0 & 57.2 & 64.4 ($\downarrow$ 7.5) & & 93.0 & 93.1 & 94.3 & 92.9 & 93.3 ($\downarrow$ 3.3) \\
        Shuffled-Asm  & 75.6 & 60.5 & 54.7 & 52.0 & 60.7 ($\downarrow$ 11.2) & & 91.7 & 91.7 & 91.5 & 90.9 & 91.5 ($\downarrow$ 5.1) \\
        Masked-Pseudo & 79.7 & 73.1 & 66.6 & 63.3 & 70.7 ($\downarrow$ 1.2) & & 95.0 & 96.8 & 96.2 & 95.0 & 95.8 ($\downarrow$ 0.8) \\ \hline
    \end{tabular}
    \label{t-attention-bias}
\end{table*}

As shown in Table~\ref{t-attention-bias}, \ASAP{} experiences substantial performance degradation across all optimization levels when the assembly view is corrupted. Notably, the average re-execution rate drops by 7.5\% and 11.2\% under the Zero-Asm and Shuffled-Asm conditions, respectively. This sensitive response confirms that our two-stage training pipeline successfully forces the LLM to extract corrective evidence from the assembly tokens rather than passively ignoring them. Furthermore, \ASAP{} maintains a highly competitive re-execution rate (70.7\% on average) under the Masked-Pseudo condition. This resilience suggests that the learned assembly tokens supply robust, source-aligned semantic cues that can guide the LLM even when pseudocode is unreliable.

\subsection{Evaluation on BringUpBench}
\label{app:bringupbench}

Following SK2Decompile~\cite{tan2025sk2decompile}, we evaluate \ASAP{} on BringUpBench~\cite{austin_bringup_bench}, a suite of nontrivial C programs with minimal dependencies. Table~\ref{t-bringupbench} reports re-execution rates across optimization levels O0--O3.

\begin{table*}[!htbp]
    \centering
    \footnotesize
    \setlength{\belowcaptionskip}{4pt}
    \caption{Re-execution rates (\%) on BringUpBench.}
    \label{t-bringupbench}
    \setlength{\tabcolsep}{3.5pt}
    \begin{tabular}{lccccc}
        \hline
        \multicolumn{1}{c}{Method} & O0 & O1 & O2 & O3 & AVG \\
        \hline
        Ghidra       & 6.6  & 5.2  & 4.5  & 2.7  & 4.8 \\
        IDA Pro      & 25.4 & 21.6 & 21.4 & 18.3 & 21.7 \\
        SK2Decompile & 34.3 & 29.1 & 24.9 & 19.7 & 27.0 \\
        \ASAP{}      & \textbf{42.6} & \textbf{34.3}
                     & \textbf{32.3} & \textbf{28.6}
                     & \textbf{34.4} \\
        \hline
    \end{tabular}
\end{table*}

Table~\ref{t-bringupbench} shows that both SK2Decompile and \ASAP{} improve re-execution rates over their respective decompiler outputs.
SK2Decompile achieves an average re-execution rate of 27.0\% using IDA Pro pseudocode, while \ASAP{} achieves 34.4\% using Ghidra pseudocode, with an absolute improvement of 7.4\%.
These results extend our evaluation beyond HumanEval and MBPP and demonstrate the effectiveness of \ASAP{} on functions drawn from nontrivial C programs.

\subsection{Additional Baseline Comparisons}
\label{app:additional_baselines}

We extend the HumanEval and MBPP evaluation with GPT-4.1-mini, GPT-5-mini, GPT-5.5, and Claude-Opus-4-8. We also include SAILR~\cite{basque2024sailr}, FAE~\cite{feng2024self}, and DecLLM~\cite{wong2025decllm}, which represent compiler-aware structuring, fine-grained assembly-source alignment, and feedback-based pseudocode repair, respectively.
Table~\ref{t-additional_baselines} reports their re-execution rates across optimization levels O0--O3.

\begin{table*}[!htbp]
    \centering
    \footnotesize
    \setlength{\belowcaptionskip}{4pt}
    \caption{Additional re-execution rates (\%) on HumanEval and MBPP.}
    \label{t-additional_baselines}
    \setlength{\tabcolsep}{3.5pt}
    \begin{tabular}{lccccccccccc}
        \hline
        \multicolumn{1}{c}{\multirow{2}{*}{Method}} & \multicolumn{5}{c}{HumanEval} & & \multicolumn{5}{c}{MBPP} \\
        \cline{2-6} \cline{8-12}
        & O0 & O1 & O2 & O3 & AVG & & O0 & O1 & O2 & O3 & AVG \\
        \hline
        \textbf{Task-Agnostic LLM} & & & & & & & & & & & \\
        \hspace{0.4em} GPT-4.1-mini    & 75.0 & 51.8 & 51.2 & 54.9 & 58.2 & & 63.5 & 55.0 & 53.2 & 46.5 & 54.5 \\
        \hspace{0.4em} GPT-5-mini      & 81.7 & 72.0 & 65.2 & 68.3 & 71.8 & & 71.2 & 67.4 & 67.8 & 63.1 & 67.4 \\
        \hspace{0.4em} GPT-5.5         & 82.9 & 73.8 & 76.2 & 72.6 & 76.4 & & 70.7 & 67.8 & 68.3 & 64.6 & 67.8 \\
        \hspace{0.4em} Claude-Opus-4-8 & 90.9 & 81.7 & 85.4 & 84.7 & 85.7 & & 76.1 & 71.9 & 72.5 & 72.2 & 73.2 \\
        \textbf{Decompilation Method} & & & & & & & & & & & \\
        \hspace{0.4em} SAILR           & 40.2 & 18.3 & 14.6 & 12.8 & 21.5 & & 54.1 & 32.8 & 28.4 & 23.0 & 34.6 \\
        \hspace{0.4em} FAE             & 67.7 & 48.8 & 45.7 & 42.1 & 51.1 & & 61.6 & 46.0 & 46.8 & 44.7 & 49.8 \\
        \hspace{0.4em} DecLLM          & 80.4 & 65.2 & 65.5 & 58.5 & 67.4 & & 74.8 & 59.5 & 58.2 & 51.3 & 60.9 \\
        \hspace{0.4em} \ASAP{}         & 87.2 & 76.8 & 72.0 & 64.6 & 75.2 & & 77.1 & 69.8 & 65.0 & 62.7 & 68.7 \\
        \hline
    \end{tabular}
\end{table*}

Claude-Opus-4-8 achieves the highest average re-execution rates among the compared methods on both datasets, reaching 85.7\% on HumanEval and 73.2\% on MBPP. GPT-5.5 and \ASAP{} obtain similar average results, with GPT-5.5 performing better on HumanEval and \ASAP{} on MBPP. These results demonstrate the strength of recent task-agnostic models as baselines for pseudocode refinement. However, public benchmarks may overlap with LLM training data, potentially inflating evaluation scores~\cite{tan2025decompilebench}.

Among the three additional decompilation baselines, DecLLM achieves the highest average re-execution rates, reaching 67.4\% on HumanEval and 60.9\% on MBPP.
\ASAP{} achieves 75.2\% and 68.7\%, respectively, corresponding to an absolute improvement of 7.8\% on both datasets.
DecLLM uses compilation and runtime feedback to repair pseudocode, while \ASAP{} incorporates source-aligned assembly representations into the refinement process.

We additionally present the HumanEval recompilation rates reported by HELIOS~\cite{achamyeleh2026helios} alongside our results at O0, O1, and O3 in Table~\ref{t-helios}. HELIOS provides hierarchical structural context for LLM-based decompilation. In its reported evaluation, structural context increases the average object-file compilability of GPT-4.1-mini from 71.4\% to 89.6\%. \ASAP{} achieves an average recompilation rate of 98.4\% across these optimization levels.

\begin{table*}[!htbp]
    \centering
    \footnotesize
    \setlength{\belowcaptionskip}{4pt}
    \caption{Recompilation rates (\%) on HumanEval.}
    \label{t-helios}
    \setlength{\tabcolsep}{3.5pt}
    \begin{tabular}{lcccc}
        \hline
        \multicolumn{1}{c}{Method} & O0 & O1 & O3 & AVG \\
        \hline
        GPT-4.1-mini
            & 85.8 & 71.0 & 57.3 & 71.4 \\
        GPT-4.1-mini + HELIOS
            & 96.6 & 83.6 & 88.6 & 89.6 \\
        \ASAP{}
            & 99.4 & 98.2 & 97.6 & 98.4 \\
        \hline
    \end{tabular}
\end{table*}

\end{document}